\documentclass[aip,apl,reprint]{revtex4-1}

\usepackage{graphicx}
\usepackage{color}
\begin{document}

\title{Epitaxial Ferromagnetic Nanoislands of Cubic GdN in Hexagonal GaN}

\author{T. F. Kent}
\affiliation{Department of Materials Science and Engineering, The Ohio State University, Columbus, Ohio 43210, USA}

\author{J. Yang}
\affiliation{Department of Materials Science and Engineering, The Ohio State University, Columbus, Ohio 43210, USA}

\author{L. Yang}
\affiliation{Department of Materials Science and Engineering, The Ohio State University, Columbus, Ohio 43210, USA}

\author{M. J. Mills}
\affiliation{Department of Materials Science and Engineering, The Ohio State University, Columbus, Ohio 43210, USA}

\author{R. C. Myers}
\affiliation{Department of Materials Science and Engineering, The Ohio State University, Columbus, Ohio 43210, USA}
\affiliation{Deparment of Electrical and Computer Engineering, The Ohio State University, Columbus, Ohio 43210, USA}

\begin{abstract}
Periodic structures of GdN particles encapsulated in a single crystalline GaN matrix were prepared by plasma assisted molecular beam epitaxy. High resolution X-ray diffractometery shows that GdN islands, with rock salt structure are epitaxially oriented to the wurtzite GaN matrix. Scanning transmission electron microscopy combined with in-situ reflection high energy electron diffraction allows for the study of island formation dynamics, which occurs after 1.2 monolayers of GdN coverage. Magnetometry reveals two ferromagnetic phases, one due to GdN particles with Curie temperature of 70K and a second, anomalous room temperature phase.
\end{abstract}

\maketitle

	 In this work, the epitaxial integration of discrete cubic GdN nanoparticles in a continuous, high crystalline quality GaN matrix is reported. Although the growth of coalesced epitaxial GdN films on III-nitrides by molecular beam epitaxy (MBE)\cite{Scarpulla,Natali} and by reactive ion sputtering on AlN\cite{Yoshitomi} has been previously reported, formation of discrete GdN islands within a continuous, epitaxial III-nitride matrix has until now, yet to be explored. Epitaxial growth of dissimilar crystal structures is in general met with many challenges\cite{Sands}, but the rare earth pnictides (RE-Pn:EuO, ErAs.) have been shown to grow well on zincblende III-V semiconductor compounds\cite{Palmstrom}. Most widely studied has been the epitaxial integration of semi-metallic ErAs in InGaAs\cite{Kadow}, which has resulted in high speed photodetector and THz source applications\cite{Griebel}. The mechanism of ErAs embedded nanoparticle growth in GaAs has been proposed\cite{Crook} to proceed first by incomplete layer formation of ErAs islands on the surface followed by epitaxial lateral overgrowth of the uncovered GaAs. 

\begin{figure}[h!]
\begin{tabular}{c}
\includegraphics[scale=0.9]{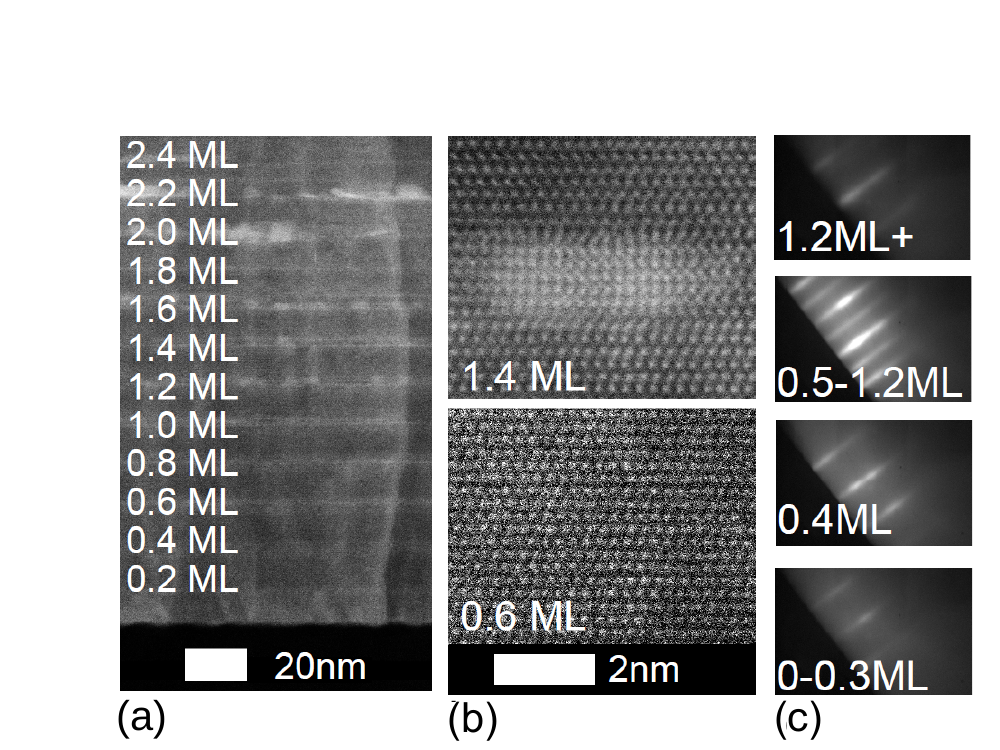}\\
\end{tabular}
\caption{\textbf{(a)} Cross sectional z-contrast HAADF STEM image of calibration sample. \textbf{(b)} Atomic resolution STEM images for selected layers in the calibration structure. The top image shows a GdN nanoparticle of clearly cubic structure surrounded by a continuous GaN matrix. The bottom atomic resolution data shows the normal wurtzite structure prior to island formation. \textbf{( c)} RHEED patterns taken during growth of GdN layer for different thicknesses showing evolution of the surface reconstruction from  1$\times$2 to 2$\times$4 reconstruction and back to the 1$\times$2 with increasing layer thickness.}
\label{calstack}
\end{figure}

\begin{figure*}
\center
\includegraphics[scale=1]{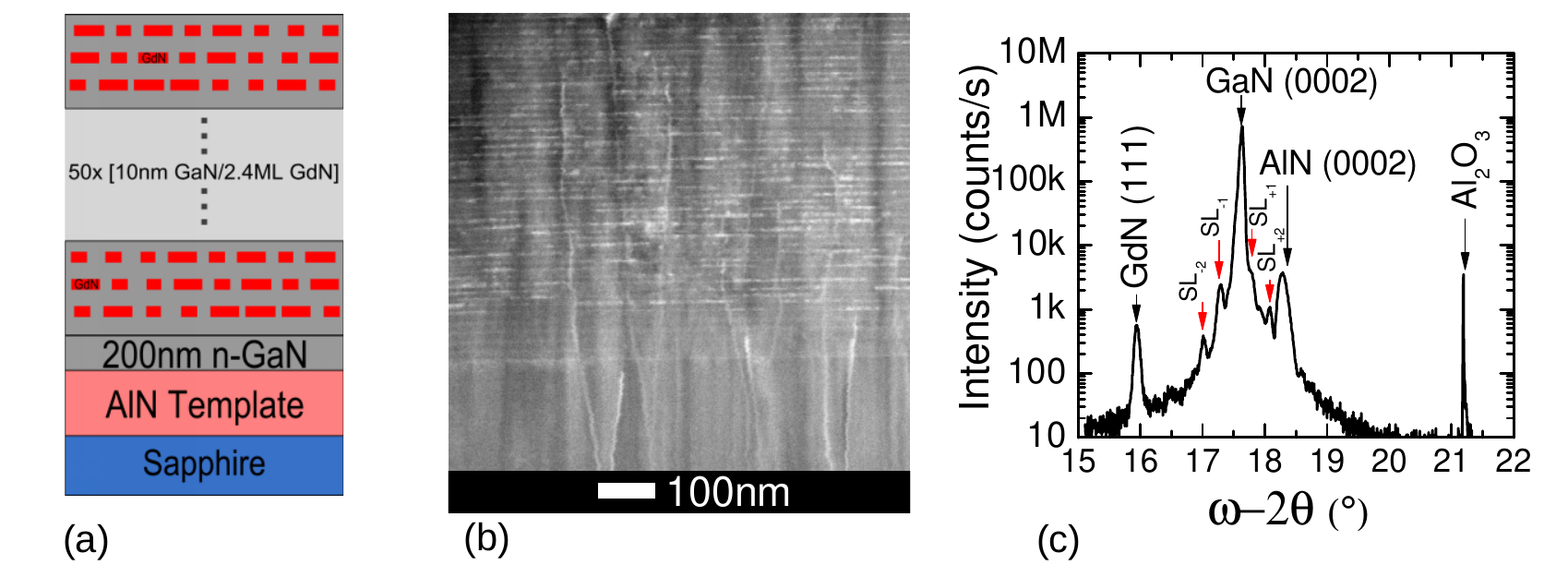}
\caption{\textbf{(a)} Structural diagram of 50 period 10nm GaN/2.4ML GdN superlattice. \textbf{(b)} Cross sectional z-contrast HAADF STEM image showing all periods of superlattice heterostructure. \textbf{(c)} High resolution X-ray diffraction $\omega$-2$\theta$ scan showing epitaxial orientation of the GdN (111) peak to the wurtzite (0002) of GaN as well as superlattice fringes, indicating precise layer thickness control.}
\label{SLstruct}
\end{figure*}

The epitaxial integration of GdN with GaN is attractive for a number of reasons. First, the dilute doping of III-Nitrides with Gd has attracted a large amount of attention in recent years initially for its promise of utilization of the intra-f-shell UV optical transitions of Gd in AlN\cite{Gruber} and subsequently for the search of a room temperature dilute magnetic semiconductor following the report of room temperature ferromagnetism in Gd:GaN\cite{Dhar, Bedoya, Davies}. Gd is attractive for its magnetic properties, possessing the most strongly correlated electronic structure of the lanthanides, with 4f ground state of spin 7/2 . Devices for semiconductor spintronics require efficient ferromagnetic spin injection and detection layers, which currently are composed of either dilute magnetic semicondutor(DMS) (GaMnAs) or metallic layers. No epitaxial spin injector is currently available in the III-nitride materials system. For the preservation of spin coherence through the device, interface and crystalline quality are key considerations. Dilute Gd:GaN, though offering the promise of room temperature ferromagnetism and realization of a nitride based DMS has proven to be a difficult material to control\cite{Roever} due to its poorly understood defect mediated mechanism of ferromagnetism. GdN, in contrast is a well understood classical ferromagnet\cite{Gambino} with T$_c$ around 70K. Furthermore, unlike most other RE-Pn, which are well established to be semimetals, thin GdN layers has been predicted\cite{Mitra,Duan,Lambrecht} to be indirect gap semiconductors, a claim which is consistent with recently reported absorption features for thin GdN films\cite{Yoshitomi}. This leads to the possibility of a controllable ferromagnetic semiconductor which can be epitaxially integrated with GaN. In addition to intriguing magnetic properties, embedded GdN nanoparticles in GaN could potentially function as carrier recombination centers, giving rise to ultrafast photoconductivity in the same fashion as RE-As particles in III-arsenides.

For the epitaxial structure of the matrix to remain single crystalline, the layer coverage of the rock salt GdN must remain incomplete, allowing for epitaxial laterall overgrowth of the surrounding matrix. This is due to the lower symmetry of the rock salt structure (Fm$\overline{3}$m) than the host (P6$_3$mc in the case of wurtzite). Complete films of GdN on wurtzite III-Nitrides have been shown to epitaxial with the relationship GdN[111]$||$GaN[0001] but containing two rotational variants due to crystal symmetry considerations, which can be observed by an off-axis $\phi$ scan in x-ray diffractometry and resulting in a polycrystalline overlayer of GaN\cite{Scarpulla}.

Samples were prepared using the technique of plasma assisted molecular beam epitaxy (PAMBE). In a Veeco GEN930 PAMBE system equipped with a Ga, Gd effusion cell, and nitrogen plasma source. To study the GdN island formation threshold, a  calibration stack consisting of increasing effective thicknesses of GdN are deposited from 0.2ML to 2.4ML in between 10nm GaN spacers on a GaN buffer layer grown on an AlN on sapphire (KYMA) template at a substrate temperature of 730C, beam equivalent pressure of 2$\times10^{-5}$Torr and III/V ratio of 2. During the period of GdN growth, the Ga shutter is closed, meaning only Gd and N are being deposited. There is, however, a residual amount of Ga present on the surface, due to growth of the GaN spacer under metal rich conditions. To analyze the onset of GdN island formation, the samples were characterized by cross-sectional, atomic resolution TEM using an FEI Titan3 80-300 Probe-Corrected Monochromated (S)TEM, as can be seen Fig \ref{calstack}a. Up to 1.2ML GdN, no change in the structure of the heavily Gd doped region is observed, however at 1.2ML, discrete clusters of highly concentrated Gd atoms appear. From the atomic resolution data shown in Fig.\ref{calstack}b, cubic particles of GdN are observed in the 1.4ML layer. From image analysis of the STEM data\cite{ImageJ}, the lattice parameter of cubic GdN in GaN is measured to be 4.8$\pm0.1\text{\AA}$ and the nanoparticle size is roughly 2.6nm x 3.6nm. After 1.2ML and up to 2.4ML, GdN particles with clearly cubic structure surrounded by a hexagonal GaN matrix can be seen with the major change with additional Gd deposition being increased lateral growth, suggesting that the height of the nanoparticle is self limited and further growth will proceed by lateral expansion of the GdN islands, which is similar to what has been observed for Er-Pn in III-As nanoparticle structures\cite{Hanson}.

\begin{figure*}
\center
\includegraphics[scale=1]{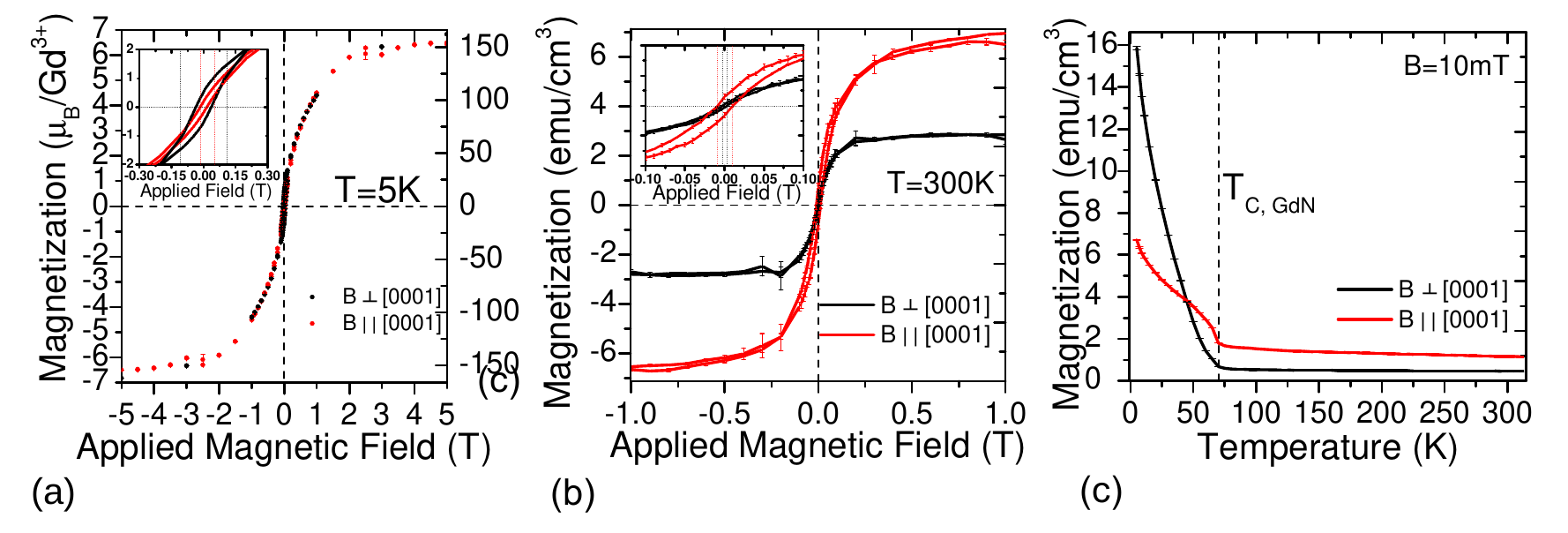}
\caption{Diamagnetic background corrected magnetization hysteresis loops for low \textbf{(a)} and high \textbf{(b)} temperatures from SQUID magnetometry for in-plane and out-of-plane film orientations relative to the applied field. Inserts show the low field data and open nature of the loops. The low temperature scan clearly shows the highly symmetric GdN magnetic phase with the expected saturation magnetization of 7$\mu$B/Gd$^{3+}$. \textbf{(c)}. Diamagnetic background corrected, constant field magnetization behavior with temperature after field cooling from 300K to 5K at 5T for in plane and out of plane orientation of the film with the applied field.}
\label{SQUID}
\end{figure*}

 During growth, the surface reconstruction was monitored using reflection high energy electron diffraction (RHEED) operating at 10kV and cathode current of 1.4A, results are shown in Fig. \ref{calstack}c. For the first 0.4ML of GdN coverage, the pattern is representative of the wurtzite Ga-face 1$\times$2 reconstruction. After 0.5ML and until 1.2ML of GdN the pattern changes to a 2$\times$4 reconstruction. Past 1.2ML of coverage, the wurtzite 1$\times$2 pattern again is visible, indicating a temporary change in the surface structure during growth of the GdN layer. 

After calibration of the GdN precipitation threshold, the heterostructure show in Fig. \ref{SLstruct}a. consisting of a GaN buffer on an AlN template on sapphire (KYMA) and a 50 period superlattice of alternating 10nm uid-GaN and 2.4ML GdN layers was prepared under identical growth conditions as the calibration structure. Cross sectional STEM images, shown in Fig.\ref{SLstruct}b, using a Technai F20 operating in HAADF imaging mode, which provides atomic number contrast, shows expected discrete GdN particles in a GaN matrix with GaN spacer thickness of 11.8$\pm$0.4nm and GdN layer thickness of 5.6$\pm$0.3nm obtained from image analysis.The structure of the sample was further characterized by high resolution x-ray diffractometry using a Bruker D8 triple axis system. Diffraction data shown in Fig. \ref{SLstruct}c. exhibits clear epitaxial orientation of the GdN [111] to the wurtzite [0001]. Also visible are superlattice fringes, indicating precise layer thickness control of the GaN spacing layers between the GdN regions. From analysis of the superlattice fringes, the GaN spacer thickness can be determined to be 10.98nm which is close to the value determined from STEM. Furthermore, from the diffraction angle, we can determine the lattice parameter of GdN to be 4.97$\text{\AA}$, which is in very good agreement with the value obtained from STEM of the nanoparticles and with values for bulk GdN\cite{Natali}.

The magnetic properties of the sample were analyzed by superconducting quantum interference device (SQUID) magnetometry using a Quantum Design MPMS XL. Results, depicted in Fig. \ref{SQUID} clearly show evidence of two distinct ferromagnetic phases in the sample. The dominant phase at low temperature (Fig. \ref{SQUID}a) can be identified as rocksalt GdN due to a saturation magnetization of nearly exactly 7$\mu_B$/Gd$^{3+}$ (158.2 emu/cm$^3$), the expected configuration for GdN. The low remanent magnetization but correct saturation of 7$\mu_{\text{B}}$ is indicative that a large fraction of Gd is paramagnetic, which is further supported by a temperature dependence containing both a mean field like behavior with a T$_{\text{c}}$ 70K but an additional 1/T contribution. Samples were characterized in both the in-plane (\textbf{$\vec{B}$}$||$GaN [0001]) and out-of-plane (\textbf{$\vec{B}$}$\perp$GaN [0001]) configuration. The low temperature phase shows very little anisotropy which is consistent with small particles of a cubic structure, which should be free from the shape anisotropy of a fully coalesced film. The coercive field, H$_{\text{c}}$ is measured to be 363Oe for the in-plane configuration and 170Oe in the out-of-plane configuration, respectively.

	Past the curie point of GdN, a second, weaker and anisotropic ferromagnetic phase persists to room temperature. This anomalous phase is hypothesized to be the result of  interaction of the Gd with local point defects in the GaN matrix and is of the same type as observed by Dhar, et. al.\cite{Dhar}. It was previously reported\cite{Davies} that ferromagnetic films of dilutely doped Gd:GaN exhibit anisotropy in their saturation magnetization between the in-plane and out of plane orientations of the film. As observed in Fig \ref{SQUID}b, the room temperature phase exhibits  anisotropy with M$_{\text{s}}$=6.84emu/cm$^3$ and H$_{\text{c}}$ = 100Oe for the out-of-plane configuration and M$_{\text{s}}$ =2.8emu/cm$^3$, H$_{\text{c}}$ = 30.6Oe for the in-plane orientation of the film.

	Measurement of the magnetization behavior with temperature, after cooling in a 5T field, is shown in Fig \ref{SQUID}c. These data reveal a sharp decrease in the magnetization with temperature up to 70K, the reported Curie point of GdN\cite{Scarpulla}. After 70K and up to the highest temperature measured, 350K, a residual amount of magnetization persists, again pointing to the possibility of an anomalous room temperature phase. In the M vs. T data, anisotropy is observed at low temperatures, which is consistent with the low field anisotropy present in the 5K hysteresis scan. For the out-of-plane configuration, a distinct knee is visible in the M vs. T scan which could be due to error in the orientation of the film as mounted in the magnetometer, causing signal from the in-plane configuration to contribute slightly to the measured magnetization. Due to the sample mounting technique employed, the out-of-plane configuration has a larger uncertainty in the absolute orientation of the film.

In summary, we have extended the growth of embedded rare earth pnictide nanoparticles in III-V semiconductors to the family of the III-nitrides. Samples show clear rocksalt structure in cross sectional TEM above a threshold value of 1.2ML GdN and x-ray diffractometry indicates epitaxial orientation of the [111] direction of GdN to the wurtzite c-axis. Magnetic characterization shows evidence of two magnetic phases, one due to the rocksalt ferromagnet GdN with Curie temperature of 70K and a second, anisotropic phase whose magnetization persists past room temperature. The room temperature phase is hypothesized to be the same type of defect mediated ferromagnetism reported in Gd:GaN and shows a prominent out-of plane easy axis\cite{Dhar,Bedoya, Davies}.

	Funding provided by the Center for Emergent Materials under NSF Award Number DMR-0820414 and the Institute for Materials Research at OSU under the Interdisciplinary Materials Research Grants (IMRG) program. Jim O'Brien of Quantum Design is thanked for valuable discussions about advanced SQUID measurement techniques.

\end{document}